\documentclass[10pt]{iopart}
\usepackage{setstack,iopams}
\usepackage{psfig,cite}
\newcommand{\PES}{P.E.S. }

\newtheorem{thm}{\bf Theorem} %
\newtheorem{prop}[thm]{\bf Proposition} %

\newcommand{\reals}{\mathbb{R}}

\newcommand{\calP}{\mathcal{P}}
\newcommand{\calM}{\mathcal{M}}

\newcommand{\hcalM}{\hat{\mathcal{M}}}

\newcommand{\hH}{\hat{H}}
\newcommand{\hT}{\hat{T}}

\newcommand{\hu}{\hat{u}}
\newcommand{\hq}{\hat{q}}
\newcommand{\hr}{\hat{r}}
\newcommand{\ha}{\hat{a}}
\newcommand{\hb}{\hat{b}}
\newcommand{\hA}{\hat{A}}
\newcommand{\hB}{\hat{B}}
\newcommand{\hpsi}{\hat{\psi}}
\newcommand{\hphi}{\hat{\phi}}

\newcommand{\ta}{\tilde{a}}
\newcommand{\tb}{\tilde{b}}

\newcommand{\ptl}{_{\scriptscriptstyle  \mathrm{PT}}}
\newcommand{\ptlp}[1]{_{{\scriptscriptstyle  \mathrm{pt}},#1}}

\newcommand{\tfrac}[2]{{\textstyle\frac{#1}{#2}}}
\newcommand{\up}[1]{^{(#1)}}

\newcommand{\sech}{\mbox{sech}}

\newcommand{\supth}{^{\mathrm{th}}}
\newcommand{\supst}{^{\mathrm{st}}}

\newcommand{\lp}{\left(}
\newcommand{\rp}{\right)}
\newcommand{\qed}{\hfill{\raise 4pt\hbox{\fbox{}}}}
\begin{document}

\title[Supersymmetry  and algebraic deformations]{Supersymmetry and algebraic deformations}

\author{D G\'omez-Ullate\dag, N Kamran \ddag$\,\;$ and R Milson\S}

\address{\dag\ Centre de Recherches Math\'ematiques, Universit\'e de
  Montr\'eal, (QC) H3C 3J7 Canada. }

\address{\ddag\ Department of Mathematics and Statistics, McGill
  University, Montr\'eal (QC) H3A 2K6 Canada.}

\address{\S\ Department of Mathematics and Statistics, Dalhousie
  University, Halifax (NS) B3H 3J5 Canada.}

\eads{\mailto{ullate@crm.umontreal.ca},
  \mailto{nkamran@math.mcgill.ca}, \mailto{milson@mathstat.dal.ca}}

\begin{abstract}
  We describe a class of algebraically solvable SUSY models by
  considering the deformation of invariant polynomial flags by means
  of the Darboux transformation.  The algebraic deformations
  corresponding to the addition of a bound state to a shape-invariant
  potential are particularly interesting.  The polynomial flags in
  question are indexed by a deformation parameter $m=1,2,\ldots$ and
  lead to new algebraically solvable models.  We illustrate these
  ideas by considering deformations of the hyperbolic P\"oschl-Teller
  potential.

\end{abstract}
 \pacs{03.65.Fd, 03.65.Ge}
 \section{Introduction}

Our purpose in this paper is to show how new classes of exactly
solvable supersymmetric quantum mechanical Hamiltonians arise in
a natural fashion from the application of the Darboux transformation to
classes of second order linear differential operators that preserve flags
of vector spaces generated by univariate polynomials. These
new Hamiltonians and their bound states have closed analytic expressions
in terms of elementary functions, and their qualitative behavior is
both natural and significant from a physical point of view.

Recall that the Darboux transformation starts from the knowledge
of a formal eigenfunction of a Schr\"odinger operator, which is
used to factorize it as a product of first-order operators.
Depending on whether this formal eigenfunction or its reciprocal
are square integrable, one obtains forward or backward Darboux
transformations, in which the supersymmetric partner Hamiltonian
is obtained by reversing the order of these factors. The principle
of our approach is to consider only those factorizations for which
the effect of the Darboux transformation on functions is to map
polynomials to polynomials. These are given a simple
characterization in our paper in terms of the starting formal
eigenfunction. We will refer to this transformation as the {\em
algebraic Darboux transformation}. The new Hamiltonians obtained
in this fashion will be exactly solvable in the precise algebraic
sense that they will also admit complete invariant flags of
polynomial subspaces.

When a parametrized family of potentials is closed with respect to
the forward Darboux transformation, it is said to be {\em
shape-invariant} \cite{gendenshtein} and in this case the
iteration of the transformation furnishes a complete description
of the spectrum and eigenfunctions. For the shape invariant
potentials, the underlying invariant flag of polynomials is the
full polynomial module. Thus, to obtain deformations of a
shape-invariant potential one must consider the two-parameter
family of backwards Darboux transformations. These were first
applied to the harmonic oscillator in \cite{mielnik}, while the
general theory was developed in \cite{deift,sukumar}. However, as
noted in \cite{levai}, the general form of the deformed potential
can only be expressed by a formal power series, or as the integral
of eigenfunctions of the original Hamiltonian -- in contrast to
the original potential, which is an elementary function, with
bound states also described by elementary functions. Our main
emphasis in this paper is to obtain examples of exactly solvable
potentials which lie outside the shape-invariant class, which can
be expressed in closed analytic form, and which have qualitative
properties that make them relevant to the description of
physically realistic situations. These are obtained by the
application of the backward Darboux transformation to shape
invariant potentials corresponding to special values of the
parameters. There is a countable infinity of algebraic backwards
transformations of a shape-invariant potential \cite{GKM3},
indexed by an integer $m$. We will show the precise manner in
which the $m^{\mbox{\scriptsize th}}$ algebraic backward
transformation deforms the invariant polynomial flag of a shape
invariant potential, and we calculate the explicit basis of the
deformed flags for the cases $m=1$ and $m=2$. As an illustrative
example we discuss in detail the algebraic deformation of the
hyperbolic P\"oschl-Teller potential.

 %This is illustrated in detail for the hyperbolic
%P\"oschl-Teller potential, following a general discussion of the
%deformations of the standard monomial flag. In general, algebraic
%deformations are indexed by an integer $m$ (the codimension). The
%algebraic exact solvability is demonstrated explicitly for the
%deformations of the hyperbolic P\"oschl-Teller potential for the
%cases $m=1$ and $m=2$.
% Classical shape-invariant operators leave invariant the standard
% polynomial flag. We will analyze the deformations of the standard flag
% that occur when applying a backwards DT (for very particular values)
% These algebraic deformations are indexed by an integer $m$, and we
% will work out how the invariant polynomial flag is deformed when $m=1$
% and $m=2$.

\section{Darboux transformations}
\subsection{The self-adjoint case.}\label{self}

Consider the Schr\"odinger operator
\begin{equation}
  \label{eq:schrodop}
  H = -\partial_{xx}+u,
\end{equation}
where $u(x),\; x\in\reals$ is continuous, real-valued and bounded from
below.  Consequently, the restriction of $H$ to a certain dense
subspace $D(H)\subset L^2(\reals)$ is a self-adjoint operator.
Consider a formal eigenfunction
 \[ H[\phi]= \lambda_0 \phi.\]
The key idea of the supersymmetric or Darboux transformation is
the fact that to every $\phi$ there corresponds a factorization of
$H$ as
\begin{equation}
  \label{eq:Hfac}
  H -\lambda_0 = A^* A,
\end{equation}
where
\begin{equation}
  \label{eq:factor-def}
  A = \partial_x-(\log\phi)_x,\quad A^*=-\partial_x-(\log\phi)_x.
\end{equation}
We shall refer to $\phi$ as the \emph{factorization function}, and to
$\lambda_0$ as the \emph{factorization energy}. The supersymmetric
partner potential is the operator defined by the
commutation of the factors
\begin{equation}\label{pot_transf}
\hH =-\partial_{xx}+ \hu=A A^*+\lambda_0,\qquad \hu= u- 2 (\log\phi)_{xx}.
\end{equation}
The transformed potential $\hu$ is continuous if and only if $\phi$ is
non-vanishing, which we assume from here on. In this way, $\hH$ is
self-adjoint and semi-bounded on some dense domain $D(\hH)$. The
operators $H$ and $\hH$ satisfy the intertwining relation
\begin{equation}
  \label{eq:intertwine}
  AH=\hH A,
\end{equation}
which  implies the following relation between the eigenfunctions
of the two operators:
\begin{equation}
  \label{eq:dxform}
H[\psi]=\lambda \psi,\quad  \hH[\hpsi] = \lambda \hpsi,\quad \hpsi=A[\psi].
\end{equation}
The spectral properties of this transformation are governed by one
of the following three possibilities\cite{sukumar,bagrov1}.
\begin{enumerate}
\item {\bf Forward transformation:} $\phi$ is square integrable (and
  since it is nodeless, it must be the ground-state wavefunction of
  $H$). The operator $A$ maps $D(H)$ onto $D(\hH)$, with a
  1-dimensional kernel.  The $n\supth$ bound state of $H$ is mapped to
  the $(n-1)\supst$ bound state of $\hH$.  Correspondingly, the
  transformed spectrum differs from the spectrum of $H$ by the removal
  of $\lambda_0$, the lowest eigenvalue.
\item {\bf Backward transformation:} $\phi^{-1}$ is square integrable.
  The operator $A$ maps the $n\supth$ bound state of $H$ to the
  $(n+1)\supst$ bound state of $\hH$.  It is one-to-one on $D(H)$, but
  not onto $D(\hH)$; the new ground state is not in the image.
  The spectrum of $\hH$ differs from that of $H$ by the addition of a
  lowest eigenvalue, namely $\lambda$, with the ground state given by
  $\phi^{-1}$.  A 2-parameter family (energy and shape parameter) of
  backward transformations exist for every $\lambda$ strictly smaller
  than the infimum of the spectrum of $H$.
\item {\bf Isospectral transformation:} neither $\phi$ nor $\phi^{-1}$
  are square integrable.  The operator $A$ defines a linear
  isomorphism from $D(H)$ to $D(\hH)$.  It transforms the $n\supth$
  bound state of $H$ to the $n\supth$ bound state of $\hH$. Two
  isospectral Darboux transformations exist for every $\lambda$
  strictly smaller than the infimum of the spectrum of $H$.
\end{enumerate}

\subsection{The general form and covariance.}
In this paragraph we will consider Darboux transformations of an
arbitrary second order operator, the general form of which is
\begin{equation}
  \label{eq:Top}
  T = p\,\partial_{zz} + q \, \partial_z + r,
\end{equation}
where we assume that $p(z)<0$ on the domain of interest.  The
above operator is related to a Schr\"odinger operator
\eref{eq:schrodop} by the change of variables
\begin{equation}
  \label{eq:varchange}
  x = \int (-p)^{-\frac{1}{2}}\, dz ,
\end{equation}
and  gauge transformation
\begin{equation}
  \label{eq:gaugexform}
  H = e^{\rho}\, T \, e^{-\rho},\qquad     \rho = \int
  \tfrac{1}{2} p^{-1}(q-\tfrac{1}{2}p_z)\, dz.
\end{equation}
The relation to the potential is given by
\begin{equation}
  \label{eq:potform}
  u = \tfrac{1}{4}p_{zz}-\tfrac12 q_{z} -\tfrac14p^{-1}\lp q-\tfrac12
  p_z)(q-\tfrac32 p_z\rp+r
\end{equation}
Since gauge transformations and changes of variable are
homomorphisms of the ring of differential operators, the Darboux
transformation is covariant with respect to these operations.

Although it is customary to work in the Schr\"odinger gauge as in
Section \ref{self}, the Darboux transformation can be defined
relative to a general coordinate and choice of gauge, as shown
below. Indeed, let
\begin{equation}
  \label{eq:Tfacfunc}
  T[\phi] = \lambda_0 \phi,
\end{equation}
be a factorization eigenfunction.
Writing
\begin{equation}
  \label{eq:phirat}
  (\log\phi)_z = \frac{a}{b},
\end{equation}
we have the following factorization of $T$:
\begin{equation}
  \label{eq:Tfac}
  T = BA+\lambda_0,
\end{equation}
where
\begin{eqnarray}
  \label{eq:Adef}
  A &=&  b\partial_z -a =b(\partial_z - (\log\phi)_z),\\
  \label{eq:Bdef}
  B &=&  \frac{p}{b} \left(\partial_z + \frac{a-b_z}{b}+\frac{q}{p}\right)
  = (p\partial_z + p(\log\phi)_z +q) b^{-1}.
\end{eqnarray}
We define the partner operator to be
\begin{equation}
  \label{eq:algpartner}
  \hT = AB+\lambda_0,
\end{equation}
and observe that the following intertwining relation holds
\begin{equation}
  \label{eq:algtwine}
  \hT A = A T.
\end{equation}
A particular case occurs when $q=1/2 \,p_z$, i.e. when the
operator is in the self-adjoint gauge. Taking
\[ b = (-p)^{1/2},\quad a= (-p)^{1/2} (\log\phi)_z,\]
we have $B=A^*$, and consequently \eref{eq:Tfac} is equivalent,
after a change of variables, to the self-adjoint factorization
\eref{eq:Hfac}. We also note that the mapping $T\mapsto \hT$ is
not canonical, but rather covariant with respect to gauge
transformations, since two different choices of the denominator in
\eref{eq:phirat}
$$(\log\phi)_z = a/b=a'/b',$$
will lead to  partner operators
 $\hT$ and $\hT'$ which are related by a gauge transformation:
\begin{equation}
  \label{eq:partnercov}
  \hskip -3em
   b^{-1}\;\hT\; b-\lambda_0 = (b')^{-1}\, \hT'\, b'-\lambda_0 = (\partial_z -
   (\log\phi)_z)(p\partial_z + q+ p(\log  \phi)_z).
\end{equation}
To effect an inverse transformation, we factorize $\hT$ with
\begin{equation}
  \label{eq:hphidef}
  \hphi = b \exp\left(-\int \left({ \frac{q}{p} +
        \frac{a}{b}}\right)dz\right)
\end{equation}
as the factorization function. A simple calculation shows that
\[ \hT[\hphi] = \lambda_0 \hphi,\]
and $\hT$ has the following form
\[\hT = p\,\partial_{zz} + \hq \,\partial_z + \hr, \] where
\begin{eqnarray}
  \hq &=& q + p_z - 2p (\log b)_z, \\
  \hr &=& -p(\log b)_{zz}+ p (\log b)_{\!z}^{\,2} - (p_z+q) (\log b)_z  \\
  \nonumber
  &&  \quad -\frac{2p a^2}{b^2} +(p_z-2q)\frac{a}{b} + q_z +
  2\lambda_0-r.
\end{eqnarray}
Thus, taking
\[ \ha = p(b_z-a) - qb,\quad \hb = pb,\]
we have  $(\log\hphi)_z = \ha/\hb$, and from \eref{eq:Adef}
\eref{eq:Bdef}  we obtain
\[ \hT = \hB \hA + \lambda_0,\]
where
\begin{eqnarray}
  \label{eq:hadef}
  \hA &=& \hb \,\partial_z - \ha = bp \left(\partial_z + \frac{a-b_z}{b}
  + \frac{q}{p}\right) =
  b^2 B,\\
  \hB &=& \frac{p}{\hb} \left(\partial_z +
    \frac{\ha-\hb_z}{\hb}+\frac{\hq}{p}\right)\\ \nonumber
  &=&\frac{1}{b}\left( \partial_z - \frac{a}{b} -
    \frac{2b_z}{b}\right) = b\left(\partial_z - \frac{a}{b}\right) b^{-2}=  A
  b^{-2}.
\end{eqnarray}
It follows that
\[ \widehat{\hT} = \hA \hB + \lambda_0 = b^2\, T \,b^{-2}.\]

\subsection{Algebraic factorizations.}
We will say that a second-order differential operator is
\emph{exactly solvable by polynomials} (P.E.S.) if it is
equivalent, by a change of variable and a gauge transformation, to
a second-order operator $T$
%(c.f. \eref{eq:Top})
that preserves an infinite flag of finite-dimensional {\em
polynomial} subspaces
\begin{equation}
  \label{eq:mflag}
  \calM_1\subset \calM_2\subset\calM_3\subset\ldots\subset\calM =
  \cup_n \calM_n.
\end{equation}
As part of this definition we include the following assumptions:
\begin{itemize}
\item[(E1)] There is a fixed polynomial codimension, which we will
  call $m$. To be more
  precise, we assume that each $\calM_n$ is an $n$-dimensional
  subspace of
  \[\calP_{n+m-1}=\langle 1,z,z^2,\ldots, z^{n+m-1}\rangle.\]
\item[(E2)] There is no spectral degeneracy. The action of $T$ is
  upper-triangular relative to a basis adapted to the above flag, and
  hence possesses an infinite list of eigenpolynomials.  We assume
  that the corresponding eigenvalues are distinct.
\end{itemize}
This definition is similar to the definition of {\em exact
solvability} introduced in \cite{turbiner1}. Of particular
interest is the subclass of \PES operators for which
$\calM_1=\reals$. If this condition holds, we will say that the
operator satisfies the \emph{algebraic ground state condition}.

We will now consider the following question:  suppose that $T$ is
a \PES operator with invariant flag \eref{eq:mflag}, what are the
conditions on a factorization function $\phi$ such that the
partner operator $\hT$ is also \PES?  To answer this question, we
define a factorization eigenfunction \eref{eq:Tfacfunc} to be of
algebraic type whenever
\[ \frac{\phi_z}{\phi}= \frac{a}{b}\]
is a rational function, where without loss of generality the
polynomials $a=a(z)$ and $b=b(z)$ are assumed to be relatively
prime. This definition is motivated by the following observation.
\begin{prop}\label{prop1}  The factorization function $\phi$ is of algebraic type, if and only
  if the operator
  \[ A=b\partial_z-a = b(\partial_z - (\log\phi)_z),\]
  transforms polynomials into polynomials.
\end{prop}
We would like to focus on those Darboux transformations that
preserve the \PES character, from here on called {\em algebraic
Darboux transformations}. According to Proposition \ref{prop1},
these are precisely the ones in which the factorization function
$\phi$ is of algebraic type. Let us describe the invariant
polynomial flag of the partner operator in each of the three cases
discussed in Section \ref{self}.

\begin{itemize}
\item[(A1)] {\bf Algebraic forward transformation:} In this case
  $\calM_1=\langle \phi\rangle$, where $\phi$ is the factorization
  function.  The operator
  \[A = \phi \partial_z - \phi_z \]
  deletes the ground state,
  and therefore the invariant flag of the partner is
  \[ \hcalM_n = A[\calM_{n+1}].
 %\quad \dim\hcalM_n=n,\quad  \codim\hcalM_n = m+\deg(\phi).
  \]
\item[(A2)] {\bf Algebraic backward transformation:} In this case, the
  new ground state $\hphi$, given by \eref{eq:hphidef}, is a rational
  function. Writing $\hphi=\ta/\tb$, we see that the first
  condition is true if and only if there exist polynomials
  $\ta=\ta(z)$, and $\tb=\tb(z)$ such that
  \begin{equation}
    \label{eq:a2cond}
    \frac{b_z}{b}-\frac{q}{p} - \frac{a}{b} =
    \frac{\ta_z}{\ta}-\frac{\tb_z}{\tb} =
    \lp \log\frac{\ta}{\tb}\rp_{\!z},
  \end{equation}
  The partner flag is given by
  \[ \hcalM_1 = \langle \ta \rangle,\quad \hcalM_{n+1} = \tb
  A[\calM_n]\oplus \langle \ta\rangle.\]

\item[(A3)] {\bf Isospectral transformation:} In this case
  both $A$ and $B$ are linear isomorphisms, and we have
  \[ \hcalM_n = A[\calM_n].\]
\end{itemize}

Of particular interest is the case where the partner operators
represent an algebraic forward/backward transformation pair, and where
both operators satisfy the algebraic ground state condition.
\begin{prop}
  \label{prop:fbpair}
  Let $T, \hT, \phi, a, b$ be as above, and suppose that $T$ satisfies
  the algebraic ground state condition. The following are equivalent:
  \begin{itemize}
  \item[\rm (i)] $T\mapsto\hT$ is an algebraic backward
    transformation, with $\hT$ satisfying the algebraic ground state condition.
  \item[\rm (ii)] $pa_z+(r-\lambda_0) b =0$.
  \item[\rm (iii)] $\hphi$ is a constant.
  \end{itemize}
\end{prop}
\emph{Proof.}  Without loss of generality, $T$ annihilates the
constants, and it is therefore of the form
\begin{equation}
  \label{eq:Tform}
  T =p\,\partial_{zz} + q\,\partial_z.
\end{equation}
The implication (iii) $\Rightarrow$ (i) follows directly from the
definitions. In order to prove the converse, suppose that (i)
holds. By \eref{eq:Bdef} \eref{eq:hphidef} we have then
\[B=\frac{p}{b}\left(\partial_z-\frac{\hphi_z}{\hphi}\right) =
\lp \frac{p\hphi}{b}\,   \partial_z\rp \hphi^{-1}.\]
Hence,
\begin{eqnarray}
  \nonumber
  T&=&BA+\lambda_0=\left(\frac{p\hphi}{b}\, \partial_z \right) \left
    ( \frac{b}{\hphi}\,
    \partial_z - \frac{a}{\hphi}\right) +\lambda_0\\
  \label{eq:Tfac2}
  &=& p\,\partial_{zz}+q \partial_z
  -\frac{p\hphi}{b}\left(\frac{a}{\hphi}\right)_{\!\!z} +\lambda_0.
\end{eqnarray}
Note that $A[1]=-a$, and therefore the first two invariant
subspaces of the partner operator $A B$ are given by
\[\langle \hphi \rangle = \hcalM_1\quad\mbox{ and }\quad \langle \hphi, a
\rangle = \hcalM_2.\] Note that $\hT$ satisfies the algebraic ground
state condition if and only if $a/\hphi$ is a polynomial.  Also, by
\eref{eq:Tform} \eref{eq:Tfac2} we have
\[ \left(\frac{a}{\hphi}\right)_z = \lambda_0 \frac{b}{p\hphi}.\]
Since (A2) implies that $\lambda_0\neq 0$,  $\hphi$ divides both
$a$ and $b$, and therefore $\hphi$ must be a constant.

Finally, let us show that (iii) is equivalent to (ii). By
\eref{eq:hphidef}, (iii) is true if and only if
\begin{equation}
  \label{eq:abclaim}
  \frac{b_z-a}{b} - \frac{q}{p}=0.
\end{equation}
By assumption,
\[ T[\phi] = p\phi_{zz}+q\phi_z=\lambda_0 \phi,\quad\mbox{and }\quad
  \phi_z = \frac{a}{b}\phi,
\]
and therefore
\[ p\lp\frac{a_z}{b} - \frac{a}{b} \frac{b_z}{b} \rp + \lp
\frac{a}{b}\rp^2 +  \frac{qa}{b} = \lambda_0,\]
or equivalently,
\[ p \,a_z-\lambda_0b = p\,a \lp   \frac{b_z-a}{b} -
\frac{q}{p}\rp, \] so that (ii) holds if and only if
\eref{eq:abclaim} does.  \qed

\section{Algebraic deformations of shape invariant potentials.}

\subsection{Shape invariance.}
Let us recall that a parameterized potential is called {\em
  shape-invariant}\cite{gendenshtein} if the forward Darboux
transformation preserves the form of the potential while altering the
value of the parameters.  In the preceding section, we pointed out
that the Darboux transformation is covariant with respect to arbitrary
changes of gauge and variable.  As a consequence, the notion of
shape-invariance makes perfect sense for general second-order
operators, and not just for operators in Schr\"odinger form.  Thus, we
will adapt the usual definition and say that a parameterized family of
\PES operators is shape-invariant if that family is closed with
respect to the forward (A1) Darboux transformation.

We now describe an important class of shape-invariant, \PES operators.
We define the {\em standard polynomial flag} to be:
\begin{equation}\label{pflag}
\reals=\calP_0\subset\calP_1 \subset\calP_2\subset\ldots\subset
\calP_n\subset\ldots,\qquad \calP_n = \langle 1,z,\ldots,z^n\rangle.
\end{equation}
The general form of a second-order operator $T$ that preserves the
standard flag is
\begin{equation}
  \label{eq:esform}
  T=p\partial_{zz} + q\partial_z+r,
\end{equation}
where $p=p(z)$ and $q=q(z)$ are, respectively, second and first
degree polynomials, and where $r$ is a constant. The family of
operators described by \eref{eq:esform} is shape invariant in the
above sense.  The ground state is given by $\phi=1$ with
$\lambda_0=r$, and   the factorization is simply
\[ T = (p\partial_z+q)\partial_z +\lambda_0.\]
The partner operator
\[ \hT = \partial_z (p\partial_z + q) +\lambda_0= p\partial_{zz} +
\hq\partial_z + \hr,\]
retains the form  \eref{eq:esform}, with
\[ \hq = p_z+q,\qquad \hr=q_z+r.\]

The corresponding non-singular potential forms --- see
\eref{eq:varchange} \eref{eq:gaugexform} \eref{eq:potform} for the
transformation formulas and \cite{GKM3} for their derivation ---
are shown in Table \ref{tab:si}.  These are the well-known,
shape-invariant potential families: the harmonic oscillator (I),
the Morse potential (II), and the hyperbolic P\"oschl-Teller
potentials (III).  Since potentials (I) and (III) are even
functions, the corresponding eigenfunctions have a well-defined
parity.  Consequently, these potentials possess two algebraic
sectors, i.e. they are exactly solvable by polynomials in two
distinct ways (see \cite{gkm} for an algebraic explanation of
potentials with multiple algebraic sectors).  The even sector
corresponds to an even gauge factor, and the odd sector to an odd
gauge factor. The parity of the algebraic sectors is reversed by a
Darboux transformation.

\begin{table}[htbp]
  \begin{center}
    \begin{tabular}{c|ccccc}
      & $\mathrm{I_e}$ & $\mathrm{I_o}$ & II &$\mathrm{III_e}$ &
      $\mathrm{III_o}$ \\ %
      \hline
      $p$ & $-4z$ & $-4z$ & $-z^2$ & $z(1-z)$ & $z(1-z)$ \\
      $q$ & $4z-2$ & $4z-6$ & $(2A-1)z-1$ &
      $(A-\tfrac32)z+1-A$&$(A-\tfrac52)z+1-A$ \\
      $r$ & $1$ & $3$& $-A^2$ &
      $-\left(\frac{A}{2}-\frac{1}{4}\right)^2$ &
      $-\left(\frac{A}{2}-\frac{3}{4}\right)^2$ \\
      $e^\rho$ & $e^{-\frac{x^2}{2}}$  & $x\,e^{-\frac{x^2}{2}}$& $
      e^{-\frac{1}{2}e^{-x}-Ax}$& $\cosh(\tfrac{x}{2})^{-A+\frac12}$ &
      $\sinh(\frac{x}{2})\cosh(\frac{x}{2})^{-A+\frac12}$  \\
      $z(x)$  & $x^2$ & $x^2$ & $e^x$ & $\cosh^2(\tfrac{x}{2})$ &
      $\cosh^2(\tfrac{x}{2})$ \\
      $U$ & $x^2$ & $x^2$ &$\tfrac14 e^{-2x} -(A+\tfrac12) e^{-x}$
      & $\tfrac14(\tfrac14-A^2)\sech^2(\tfrac{x}{2})$
      & $\tfrac14(\tfrac14-A^2)\sech^2(\tfrac{x}{2})$
    \end{tabular}
    \caption{Shape-invariant potentials on the line}
    \label{tab:si}
  \end{center}
\end{table}

\subsection{Deformations of the standard flag.}
Let $a=a(z), b=b(z)$ be relatively prime polymomials, and let
$g=g(z)$ be a polynomial that divides $a_z$, $b$ and $b_z-a$.
Consider the differential operators
\begin{equation}
  \label{eq:defpair}
  B = g^{-1}\partial_z,\quad A = b\partial_z-a,
\end{equation}
and note that, by assumption, the second order operator
\begin{equation}
  \label{eq:defpaircond}
  T=BA = p\partial_{zz} + q\partial_z - a_z\, g^{-1}
\end{equation}
where
\[ p = bg^{-1},\qquad q =  (b_z-a) g^{-1},\]
has polynomial coefficients.  We will say that $A, B$ constitute a
\emph{deformation pair} of order $m=\deg(g)$ if $T$ leaves
invariant $\calM_n = \calP_{n-1}$ for all $n$, i.e. if $T$ leaves
invariant the standard polynomial flag. Deformation pairs are of
interest because they provide non-trivial examples of exactly
solvable Hamiltonians outside the shape-invariant class. As usual,
we define a partner operator
\begin{equation}
  \label{eq:hqform}
  \hT=AB = p\partial_{zz} + \hq\partial_z,\qquad \hq = -(b g_z+a)g^{-1},
\end{equation}
and a partner flag
\[ \reals=\hcalM_1 \subset \hcalM_2 \subset \hcalM_3 \subset
\ldots,\qquad \hcalM_{n} =
A[\calP_{n-2}]\oplus\reals,\]
which we will refer to as a \emph{deformation} of the standard
polynomial flag.
\begin{prop}
  \label{prop:defcodim}
  Every $\hcalM_n$ is a codimension $m$ subspace of $\calP_{n+m-1}$.
\end{prop}
\emph{Proof.} By assumption, $a_z g^{-1}$ is constant, and
$\deg(bg^{-1}) \leq 2$. This implies that $\deg(a) = m+1$ and that
$\deg(b)\leq m+2$.  Therefore, $\hcalM_n\subset\calP_{n+m-1}$. Since
$a, b$ are relatively prime, $A$ does not annihilate any polynomial,
and hence $\dim\hcalM_n=n$.  \qed

\begin{prop}
  The partner operator $\hT$ is \PES
\end{prop}
\emph{Proof.}  The operator $\hT$ preserves the deformed flag because
of the intertwining relation
\[ \hT A = A T,\]
and because $\hT$ annihilates $\reals=\hcalM_1$ by construction.
Condition (E1) is true by the preceding proposition.  We noted
above that $A$ and hence that $T$ does not annihilate any
polynomial.  Hence $0$ is not an eigenvalue of $T$, which proves
condition (E2).\qed

Let us also note that a deformation pair satisfies the conditions
of Proposition~\ref{prop:fbpair}, and in particular $T\mapsto\hT$
is corresponds to an algebraic backward Darboux transformation.

\begin{prop}
The deformed subspaces $\hcalM_n$ can be characterized as the subspace of
$\calP_{n+m-1}$ consisting of all polynomials $f=f(z)$ such that $g$
divides $f_z$.
\end{prop}
\emph{Proof.}
First, note that for
\[f=A[h], \quad h\in \calM_{n-1}=\calP_{n-2},\; f\in \hcalM_{n}\]
we have that $g^{-1} f_z = T[h],$ which proves that $f_z$ is
divisible by $g$.  In order to prove the converse, let us note
that the subspace of all $f\in \calP_{n+m-1}$ such that $f_z$ is
divisible by $g$ is $n$-dimensional.  However, $\dim\hcalM_n=n$ by
Proposition \ref{prop:defcodim}, which proves the claim.  \qed

We will now show an explicit basis of $\hcalM_n=n$ in the cases
$m=1$ and $m=2$. In the first instance, since the subspaces
$\calP_n$ are invariant with respect to translations, we may
without loss of generality assume that $g(z)=z$.  By
\eref{eq:defpaircond} necessarily, $b=pz$, where
$p=p_2z^2+p_1z+p_0$ is a polynomial of degree 2 or less. In order
for $g$ to divide $a_z$ and $b_z-a$ we must have
\begin{equation}
  \label{eq:m1Aform}
  A = (p_2 z^2 + p_1 z + p_0) z \partial_z - (a_2 z^2 +p_0),
\end{equation}
where $a_2, p_0, p_1, p_2$ are arbitrary real numbers.  Let us also
assume that $a_2\neq 0$ and that $a_2/p_2$ is not a positive integer.
If these generic conditions hold, then the subspaces of the partner
flag are given by
\[ A[\calP_{n-2}]\oplus\reals=\hcalM_{n} = \langle 1, z^2, z^3,\ldots,
z^{n} \rangle \] The above monomial-generated subspace is
exceptional in that it admits a seven dimensional vector space of
second-order operators that preserve it\cite{turbiner2}, and
consequently can be used to construct novel instances of exactly
solvable and quasi-exactly solvable operators \cite{GKM3,gkm}.

Turning to the case $m=2$, we limit our discussion to the generic case
of $g(z)$ with distinct roots.  By scaling and translating $z$, as
necessary, we may assume, without loss of generality, that $g=z^2-1$.
By \eref{eq:defpaircond}, in order for
$g$ to divide $a_z$ and $b_z-a$ we must have
\begin{equation}
  \label{eq:m2Aform}
   A = (p_2 z^2 + p_1 z + p_0) (z^2-1) \partial_z +(p_2+p_0)(z^3-3z)
   -2 p_1,
\end{equation}
where $p_0, p_1, p_2$ are arbitrary real numbers.
Let us also assume that $p_2+p_0\neq 0$ and that $-p_0/p_2$ is not a
positive integer.
If these generic conditions hold, then the subspaces of the partner
flag are given by
\[ A[\calP_{n-2}]\oplus\reals=\hcalM_{n} = \langle 1, \pi_3(z),\pi_4(z),\ldots
\pi_{n+1}(z)\rangle, \]
where
\begin{equation}
  \label{eq:pidef}
  \pi_{2k+1}(z) = z^{2k+1} - (2k+1)z,\qquad \pi_{2k}(z) = z^{2k} - k
  z^2.
\end{equation}
The above polynomials $\pi=\pi(z)$ have the property that $\pi_z$
is divisible by $z^2-1$.  The resulting polynomial subspaces
$\calM_{n}$ are preserved by the following second-order operators:
\begin{eqnarray*}
  T_3 &=& z^3 \partial_{zz} + \left((1-n)z^2-5+n-\frac{4}{z^2-1}\right)
  \partial_z ,\\
  T_2 &=& (z^2-1)\partial_{zz} - 2z \partial_z, \\
  T_1 &=& z\partial_{zz} - 2\left( 1+
    \frac{2}{z^2-1}\right)\partial_z,\\
  T_0 &=& \partial_{zz}+ \left( z-\frac{4z}{z^2-1}\right)
  \partial_z.
\end{eqnarray*}

\subsection{Algebraic deformations of the hyperbolic P\"oschl-Teller potential}

\begin{figure}[htbp]
  \begin{center}
    \noindent\psfig{figure=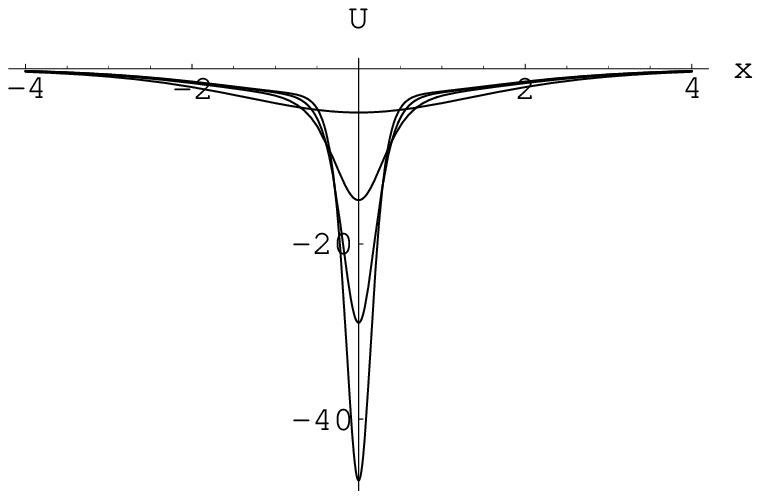,width=4in}
    \caption{ Algebraic deformations $U\up{m}\ptl(x)$  of the hyperbolic
      P\"oschl-Teller potential \eref{eq:modptell}, with $\alpha=4$ and $m=0,1,2$ and $3$.}
    \label{fig:pt}
  \end{center}
\end{figure}

The hyperbolic P\"oschl-Teller potential\cite{ptell}, which
includes the class of reflectionless 1-soliton
potentials\cite{matveev}, has the form
\begin{equation}\label{PT}
U\ptl(x)=\tfrac{1}{4}\lp \tfrac{1}{4}- \alpha^2\rp\,
\sech^2(\tfrac{x}{2}).
\end{equation}
The general solution\cite[Sec.~2.9]{bateman} of the corresponding
Schr\"odinger equation
\[ H\ptl(\phi)=-\phi_{xx} + U\ptl\,\phi=-k^2 \phi\]
can be given as
\begin{eqnarray*}
  \hskip -5em
  \phi\ptl(x;k,C_0,C_1) &=&
  \cosh(\tfrac{x}{2})^{\frac{1}{2}-\alpha}\Big\{ C_0\,\,
  {}_2F_1(-\tfrac{\alpha}{2}+\tfrac{1}{4}+k,-\tfrac{\alpha}{2}+\tfrac{1}{4}-
  k,\tfrac{1}{2};-\sinh^2(\tfrac{x}{2}))\\
  && +  C_1\, \sinh(\tfrac{x}{2})\,
  {}_2F_1(-\tfrac{\alpha}{2}+\tfrac{3}{4}+k,-\tfrac{\alpha}{2}+\tfrac{3}{4}-
  k, \tfrac{3}{2};-\sinh^2(\tfrac{x}{2})) \Big\},
\end{eqnarray*}
where ${}_2F_1(a,b,c;z)$ also denotes the analytic continuation
of the hypergeometric function to $\mathrm{Re}(z)<0$.  For
$\alpha>1/2$ , the potential \eref{PT} has $\lceil \alpha-\tfrac12 \rceil$
bound states
\[\psi\ptlp{i}(x),\quad 0\leq i<\alpha-\tfrac12.\]
The even bound states are given by \cite{GKM3}
\begin{eqnarray}
  \label{eq:pt-even}
   \psi\ptlp{2j}(x)&\propto&
   \phi\ptl(x;\tfrac{\alpha}{2}-j-\tfrac{1}{4},1,0) \\
   \nonumber
   &\propto &
   \cosh(\tfrac{x}{2})^{\frac{1}{2}-\alpha}\,
   P_j^{(-\frac{1}{2},-\alpha)}(\cosh x) .
\end{eqnarray}
The odd ones are given by
\begin{eqnarray}
  \label{eq:pt-odd}
  \hskip -3em
  \psi\ptlp{2j+1}(x) &\propto&
  \phi\ptl(x;\tfrac{\alpha}{2}-j-\tfrac{3}{4},0,1)\\
  \nonumber
  &\propto&
  \sinh(\tfrac{x}{2})\cosh(\tfrac{x}{2})^{\frac{1}{2}-\alpha}\,
  P_j^{(\frac{1}{2},-\alpha)}(\cosh x),
\end{eqnarray}
where $P_j^{(a,b)}(z)$ are the Jacobi polynomials. We  focus on
deformations of potentials with bound states only, i.e. we must
take $\alpha>\tfrac{1}{2}$.  In order to have a well defined backwards
Darboux transform of the hyperbolic P\"oschl-Teller potential, we
must consider the solutions $\phi\ptl$ which correspond to an
energy below the spectral minimum and are nowhere vanishing.
Since the spectral minimum is $-(\tfrac{1}{2}-\alpha)^2$ we take
$|k|>\tfrac{1}{2}-\alpha$.  Now, the two-parameter family of backward
Darboux transformations is given by the transformation functions:
\begin{equation}
  \label{eq:pt+-}
  \phi\ptl(x;k,1,t),\quad |t|\leq
  2\,\frac{\Gamma(\frac{3}{4}+k-\frac{\alpha}{2})
  \Gamma(\frac{3}{4}+k+\frac{\alpha}{2})}{
    \Gamma(\frac{1}{4}+k-\frac{\alpha}{2})
    \Gamma(\frac{1}{4}+k+\frac{\alpha}{2})},
\end{equation}
with the extreme values of the shape parameter $t$ corresponding
to an isospectral transformation\cite{GKM3}. In general, for an
arbitrary value of the energy parameter $k$ the partner potential
will be defined formally by a power series. However, for specific
values of $k$ the log-derivative of $\phi\ptl$ will be a
polynomial in $\cosh x$ thus giving rise to an algebraic
deformation. It can be shown that these algebraic deformations
occur precisely for the following countable subset of
\eref{eq:pt+-}
\begin{eqnarray*}
  \phi\up{m}\ptl(x) &=&\phi\ptl(x,-\tfrac{\alpha}{2}-\tfrac{1}{4} - m,1,0)\\
  &\propto&
  \cosh(\tfrac{x}{2})^{\tfrac{1}{2}+\alpha}P_m^{(-\frac12,\alpha)}(\cosh(x)),
\end{eqnarray*}
The resulting deformed potentials, as given by \eref{pot_transf},
have the form
\begin{equation}
  \label{eq:modptell}
  U\up{m}\ptl(x) =
  -\tfrac{1}{4} (\alpha+\tfrac12)(\alpha+\tfrac32)\, \sech^2(\tfrac{x}{2}) - 2\,
  \lp \log
  P_m\up{-\frac{1}{2},\alpha}(\cosh x)\rp_{xx},
\end{equation}
and have been plotted in Figure \ref{fig:pt}.

More specifically, the first ($m=1$) and second ($m=2$)
deformations have the following forms:
\[
\hskip -3em U\up{1}\ptl(x)= U\up{0}\ptl(x) + \frac{2\alpha+1}{z_1}
-\frac{4(\alpha+1)}{z_1^{\,2}},
\]
where
\[
  z_1= \tfrac12 ((2\alpha+3)\cosh x -(2\alpha+1)),
\]
and
\[
\hskip -3em U\up{2}\ptl(x)= U\up{0}\ptl(x) +\frac{(2\alpha+1)\lp \beta
  (z_2^{\,3}+3z_2)-2z_2^{\,2}-2\rp-8 }{(z_2^{\,2}-1)^2},
\]
where
\[
  z_2= \tfrac{1}{4}\beta \,((2\alpha+7) \cosh x-(2\alpha+1)),\qquad
  \beta = \sqrt{\frac{2\alpha+5}{3\alpha+6}}.
\]
The algebraic backwards Darboux transformation corresponds to the
first order operator
\begin{eqnarray*}
  A\ptl\up{m} &=& \partial_x -(\log \phi\up{m}\ptl)_x\\
  &=& \partial_x - \tfrac12(m+\alpha+\tfrac12)\sinh x
  \;\frac{P_{m-1}^{(\frac12,\alpha+1)}(\cosh
  x)}{P_m^{(-\frac12,\alpha)}(\cosh x)}.
\end{eqnarray*}
Both the undeformed and the deformed potentials are even
functions, and consequently the corresponding Hamiltonians leave
invariant the spaces of odd and even functions.  The Darboux
transformation changes parity.  In particular, the deformed even
sector is the $A$-image of the undeformed odd sector.

We now determine explicitly a basis of the invariant flag
corresponding to the even sector of the first and second
deformation. To do so, we switch to the algebraic variable and
perform a change of gauge so that the undeformed, odd algebraic
sector is isomorphic to the standard polynomial flag (c.f.  Case
$\mathrm{III_o}$ of Table \ref{tab:si}).
\begin{eqnarray}
  \label{eq:ptlT}
  T &=& z(1-z) \partial_{zz} +((\alpha-\tfrac52)z
  +1-\alpha)\partial_z-\left(\tfrac{\alpha}{2}-\tfrac{3}{4}\right)^2 ,\\ \nonumber
  T&=&e^{-\rho}\, H\ptl \, e^\rho,\qquad e^\rho =
  \sinh(\tfrac{x}{2})\cosh(\tfrac{x}{2})^{\frac12-\alpha},\\ \nonumber
  z&=&\cosh^2(\tfrac{x}{2}) = \tfrac12(\cosh x+1).
\end{eqnarray}
In the algebraic gauge, the factorization functions for the backward
transformations are given by
\[ \phi = e^{-\rho} \phi\up{m}\ptl = z^\alpha (z-1)^{-\frac12}
P_m^{(-\frac12,\alpha)}(2z-1) \]
This factorization function is of algebraic type with
\begin{eqnarray*}
  \nonumber
   &&\hskip -3em
   (\log\phi)_z = \frac{\alpha}{z} +  \frac{\frac12}{1-z}+  (\tfrac12
   +\alpha+m) \frac{
    P_{m-1}^{(\frac12,\alpha+1)}(2z-1)}{P_m^{(-\frac12,\alpha)}(2z-1)}
  =\frac{a}{b},\\
   &&  \hskip -3em
   a= ((\tfrac12-\alpha)z+\alpha)P_m^{(-\frac12,\alpha)}(2z-1)
  +(\tfrac12 +\alpha+m)z(1-z)
  P_{m-1}^{(\frac12,\alpha+1)}(2z-1),\\
  &&\hskip -3em
  b= z(1-z)P_m^{(-\frac12,\alpha)}(2z-1).
\end{eqnarray*}
A direct calculation shows that the above functions satisfy
condition \eref{eq:abclaim}, and therefore $\hphi=1$. The
operators
\[ A = b\partial_z-a,\quad B = g^{-1} \partial_z,\]
where
\[ g = P_m^{(-\frac12,\alpha)}(2z-1), \]
constitute a deformation pair of order $m$. Let us consider the
cases $m=1$ and $m=2$ in more detail.  For $m=1$ we have
\begin{eqnarray*}
  A &=& \frac{z_1 (1-z_1)(z_1+2(\alpha+1))}{2(2\alpha+3)}\,\partial_{z_1}+
  \frac{(2\alpha+1)z_1^2-4(\alpha+1)}{4(2\alpha+3)}\\
  B &=&\frac{2}{z_1}\, \partial_z,\\
  \hT &=& z(1-z)\partial_{zz} + \lp \frac{z_1}{2}
  -\frac{4(\alpha+1)}{(2\alpha+3)z_1}+\frac{2\alpha+1}{2\alpha+3}\rp
  \partial_z\\
  z_1&=&2P_1^{(-\frac12,\alpha)}(2z-1)= (2\alpha+3) z-2\alpha-2
\end{eqnarray*}
The operator $A=b(z_1)\partial_{z_1}-a(z_1)$, relative to the $z_1$
variable, is of the form \eref{eq:m1Aform}, and hence, the partner
operator $\hT$ is \PES with invariant subspaces
\[ \hcalM_{n+1} = \langle 1,z_1^{\,2},z_1^{\,3},\ldots,z_1^{\, n}\rangle.\]
For $m=2$ we set
\[   z_2=\tfrac{1}{2}\beta\,\lp (2\alpha+7)z-(2\alpha+4)\rp ,\]
so that
\[ P_2^{(-\frac12,\alpha)}(2z-1) = \frac{3(\alpha+2)}{2(2\alpha+7)}\,
(z_2^{\,2}-1).\]
Consequently,
\begin{eqnarray*}
  &&\hskip -5em
  A = \lp \frac{-6\beta
  z_2^{\,2}}{2\alpha+5}-\frac{3(2\alpha+1)z_2}{3\alpha+6} +3\beta\rp
  \frac{3(\alpha+2)^2 (z_2^{\,2}-1)}{2 (2\alpha+7)^2}\;\partial_{z_2} \\
  &&  + \frac{3(\alpha+2)}{(2\alpha+7)^2} \lp
   -\frac{(2\alpha+3)}{2\beta}(z_2^{\,3}-3z_2) -2\alpha-1\rp ,\\
  && \hskip -5em
  B = \frac{2(2\alpha+7)}{3(\alpha+2)(z_2^{\,2}-1)}\, \partial_z,\\
  &&\hskip -5em
  \hT = z(1-z)\partial_{zz} + \left[ \frac{3\beta(\alpha+2)}{2\alpha+5}
  \lp z_2 - \frac{4(2\alpha+3)}{2\alpha+7} \frac{z_2}{z_2^{\,2}-1}\rp
  + \frac{2(2\alpha+1)(z_2^{\,2}+1)}{(2\alpha+7)(z_2^{\,2}-1)}\right]
  \;
  \partial_z
\end{eqnarray*}
 In the same manner, it can be seen from
\eref{eq:m2Aform} that the partner operator $\hT$ is \PES with
invariant subspaces (c.f. \eref{eq:pidef})
\[ \hcalM_n = \langle 1,\pi_3(z_2),\pi_4(z_2),\ldots\pi_{n+1}(z_2)\rangle.\]

\section{Discussion}

In this paper we have analyzed the connection between the Darboux
transformations and exact solvability by polynomials, i.e. the
fact that a certain Hamiltonian operator after a change of
variables and a gauge transformation admits an infinite flag of
invariant polynomial subspaces. Since operator composition and
factorization are covariant with respect to changes of variables
and gauge transformations, concepts like the Darboux
transformation or shape-invariance are also covariant. It is
customary for physical applications to work in the Schr\"odinger
gauge and the physicial variable $x$, but for the purposes of
analyzing invariant flags it is more convenient to work in the
algebraic variable $z$ and in the algebraic gauge, the one in
which the operator has polynomial eigenfunctions.

A general backwards transformation on a shape invariant potential
will lead to a transformed potential whose eigenfunctions are no
longer elementary functions. We have discussed the special class
of {\em algebraic} Darboux transformations of shape-invariant
potentials, i.e. those that preserve the exact solvability by
polynomials, showing also how the polynomial flag is deformed by
the action of the Darboux transformation.

In this paper we placed our emphasis on the action of the Darboux
transformation on the invariant flag of subspaces, rather than on
the potential. In fact, many different potentials have the same
invariant flag, e.g. all the shape-invariant potentials preserve
the standard polynomial flag. We have analyzed the deformations of
the P\"oschl-Teller potential, but similar deformations exist for
other shape-invariant forms.

\ack The research of DGU is supported in part by a CRM-ISM
Postdoctoral Fellowship and the Spanish Ministry of Education
under grant EX2002-0176. The research of NK and RM is supported by
the Natural Sciences and Engineering Research Council of Canada.
NK and DGU would like to acknowledge partial financial support
from the project BFM2002-02646 of the Direcci\'on General de
Investigaci\'on.


\vskip 1cm

\begin{thebibliography}{99}
\bibitem{gendenshtein} Gendenshtein L 1983 {\it JETP Lett} {\bf 38}  356.
\bibitem{mielnik}  Mielnik B 1984 {\it J. Math. Phys.}  {\bf 25}
  3387.
\bibitem{deift} Deift P and Trubowitz E 1979 {\it Duke Math J.} {\bf
    45}, 267.
\bibitem{sukumar} Sukumar CV 1985 {\it J. Phys. A} {\bf 18} 2917.
\bibitem{bagrov1}
Bagrov V G and Samsonov B F 1995 {\it Theoret. and Math. Phys.}
{\bf  104} 1051.
\bibitem{levai} L\'evai G, Baye D and  Sparenberg J-M 1997 {\it
    J. Phys. A}  {\bf 30} 8257.

\bibitem{GKM3} G\'omez-Ullate D, Kamran N and Milson R 2004 {\it
J. Phys. A} {\bf 37} 1789.

\bibitem{turbiner1}
Turbiner A V 1988 {\it Commun. Math. Phys.} {\bf 118} 467.

\bibitem{sparenberg}Sparenberg J-M and Baye D 1995 {\it J. Phys. A}
    {\bf 28} 5079.

\bibitem{gkm} G\'omez-Ullate D,  Kamran N  and Milson R, {\em Preprint} nlin.SI/0401030.
\bibitem{turbiner2}
Post G and Turbiner A V 1995 {\it Russian J. Math. Phys.} {\bf 3}
113.

\bibitem{ko}
Kamran N and Olver P J 1990 {\it J. Math. Anal. Appl.} {\bf 145}
342.
\bibitem{gko}
Gonz\'alez-Lopez A,  Kamran N and  Olver P J 1993 {\it Commun.
Math. Phys.} {\bf 153} 117.

\bibitem{ptell}
P\"oschl G and Teller E 1933 {\it Z. Physik} {\bf 83} 143.

\bibitem{matveev}  Matveev V and Salle M A 1991 {\it Darboux transformations and
solitons}, Springer Series in Nonlinear Dynamics (Berlin:Springer)
\bibitem{bateman}
Erd\'elyi A et al. 1953 {\em Higher Transcendental Functions, Vol.
I}, (New York:McGraw-Hill).
    %
%\bibitem{darboux} Darboux G, \emph{Th\'eorie G\'en\'erale des
%    Surfaces}, vol. II, Gauthier-Villars, 1888.
%\bibitem{jacobi} Jacobi CG, 1837 \emph{J. Reine Angew. Math.} {\bf 17}, 68.
%\bibitem{schrodinger}
%Schr\"odinger E 1941 {\it Proc. Roy. Irish Acad.} {\bf 47} A, 53
%({\it Preprint} physics/9910003).
%\bibitem{infeld-hull}
%Infeld L and Hull T E 1951 {\it Rev. Mod. Phys.} {\bf 23} 21.
%\bibitem{cooper}
%Cooper F, Khare A and Sukhatme U 1995 {\it Phys. Rep.} {\bf 251}
%267.
%
%\bibitem{gesztesy} Gesztesy F, Simon B and Teschl G 1996 {\it
%    J. d'Analyse Math.} {\bf 70}, 267
%\bibitem{calogero} Calogero F and Degasperis A  1982 {\it  Spectral
%    transform and solitons I}, Studies in Mathematics and its
%    Applications (New York:Elsevier).
%\bibitem{milson} Milson R 1998 {\it Internat. J. Theoret. Phys.}  {\bf 37} 1735.
%\bibitem{ggr} G\'omez-Ullate D, Gonz\'alez-L\'opez A and
%Rodr\'{\i}guez M A 2000 {\em J. Phys. A} {\bf 33} 7305.
%\bibitem{morse}
%Morse P M 1929 {\it Phys. Rev.} {\bf 57} 57.
%
%\bibitem{baye} Baye D, Sparenberg J-M and L\'evai G 1997 {\em Inverse
%    and Algebraic Quantum Scattering Theory (Lecture notes in Physics
%    488)} ed B Apagyi, G Endr\'edi and P L\'evay (Berlin: Springer) p
%    295
%
%\bibitem{gt} Gonz\'alez-L\'opez A  and Tanaka T,
%  hep-th/0307094.
%\bibitem{shifman} Shifman  M 1989 {\em Int. J. Modern Phys. A} {\bf 4} 3311.
%\bibitem{schminke}Schminke U W  1978, {\it Proc. Roy. Soc. Edinburgh
%     Sec. A} {\bf 80}, 67.
%\bibitem{samsonov1} Samsonov B F 1999 {\it Phys. Lett. A}{\bf 263} 274.

% \bibitem{crum}
% Crum M M 1955 {\it Quart. J. Math.}  {\bf 6}  121.
% (physics/9908019).
% \bibitem{krein} Krein M G 1957
%   {\it Dokl. Akad. Nauk SSSR (N.S.)}  {\bf 113}  970.
%\bibitem{arrigo} Arrigo D J and Hickling F 2003 {\it J. Phys. A.} {\bf
%    36} 1615.

%\bibitem{dubov} Dubov S Y,  Eleonskii V M  and Kulagin N E 1992,
%  {\it Sov. Phys. JETP} {\bf 75} 446.
%\bibitem{bagrov2} Bagrov V G and Samsonov B F 1997 {\it Pramana J. Phys.} {\bf
%    49} 563.
%



\end{thebibliography}
\end{document}